\documentclass[twocolumn,preprintnumbers,amsmath,amssymb]{revtex4}

\usepackage{color}
\usepackage{graphicx}
\usepackage{dcolumn}
\usepackage{bm}
\usepackage{url}

\newcommand{\eq}[1]{Eq.~(\ref{#1})}

\begin{document}

\preprint{CERN-PH-TH-2014-124}

\title{Can the differences in the determinations of $V_{ub}$ and $V_{cb}$ be explained by New Physics?}
\author{Andreas Crivellin}
\affiliation{CERN Theory Division, CH-1211 Geneva 23, Switzerland}%
\author{Stefan Pokorski}
\affiliation{Institute of Theoretical Physics, Department of Physics,
  University of Warsaw}
\date{\today}
\begin{abstract}
The precise determination of the CKM elements $V_{cb}$ and $V_{ub}$ is crucial for any new physics analysis in the flavour sector. Their values can be determined from several tree-level decays: $V_{cb}$ can be extracted from $B\to D\ell\nu$ and $B\to D^*\ell\nu$ while $V_{ub}$ can be obtained from $B\to \pi\ell\nu$, $B\to \rho\ell\nu$ and $B\to\tau\nu$. In addition, for both $V_{cb }$ and $V_{ub}$ an inclusive determination is available. There is a long lasting discrepancy between the inclusive and exclusive determinations which recently even increased for $V_{cb}$ above the $3\;\sigma$ level. 
In this article we study the possible effect of new physics on the inclusive and exclusive determination of $V_{cb}$ and $V_{ub}$ in a model independent way. We find that there is only one operator corresponding to a modified $W$ coupling which can achieve this. However, respecting $SU(2)$ gauge invariance at the high scale this would lead to very large violations of the $Z$ to $b\bar b$ coupling not compatible with experiment. Therefore, we conclude that a new physics explanation of the difference between the inclusive and exclusive determination of $V_{cb}$ and $V_{ub}$ is currently ruled out. Therefore, the discrepancies must be due underestimated uncertainties in the theoretical and/or the experimental analysis.
  \end{abstract}
\pacs{12.15.Hh, 13.20.He}
\maketitle

\section{\label{sec:level1}Introduction}

The precise determination of the CKM elements $V_{cb}$ and $V_{ub}$ is crucial for any analysis of new physics (NP) in the quark flavour sector (see for example~\cite{Beringer:1900zz,Ricciardi:2013cda,Ricciardi:2014iga} for a review). In the standard model (SM) with its gauge group $\rm{SU(3)_C \times SU(2)_L\times U(1)_Y}$ the tree-level $W$ coupling is purely left-handed and all charged-current processes are mediated by the $W$ boson only. This property is used to extract CKM elements from tree-level decays, i.e. from exclusive leptonic and semi-leptonic decays as well as from the inclusive processes. While these processes (since they appear at tree-level) are in most analysis assumed to be free of NP. However, physics beyond the SM can in principle affect the determination $V_{cb}$ and $V_{ub}$. Furthermore, the impact of NP on the exclusive and inclusive determination is in general different.

The current situation concerning the determination of $V_{cb}$ and $V_{ub}$ is the following: For the CKM element $V_{cb}$ there has been a persistent discrepancy of slightly more than $2\,\sigma$ between the exclusive and inclusive determination for many years. Recently, new results for the $B\to D^*\ell\nu$ form-factors have been obtained on the lattice \cite{Bailey:2014tva} increasing the discrepancy between the inclusive and exclusive determination above the $3\,\sigma$ level. Concerning $V_{ub}$, the difference between the inclusive and exclusive determination increased some time ago because of the NNLO QCD corrections~\cite{Greub:2009sv}.

In detail, the current situation for the determination of $V_{cb}$ is the following:
\begin{align}
|V_{cb}|&=(4.242\pm 0.086)\times 10^{-2}\;\;({\rm inclusive})\,,\\
|V_{cb}|&=(3.904\pm 0.075)\times 10^{-2}\;\;(B\to D^*\ell\nu)\,,\\
|V_{cb}|&=(3.850\pm 0.191)\times 10^{-2}\;\;(B\to D\ell\nu)\,.
\end{align}
Here we added all errors in quadrature. The inclusive determination in taken from Ref.~\cite{Gambino:2013rza}. For the experimental input for $B\to D^*\ell\nu$ we used the HFAG average \cite{Amhis:2012bh} and for the form-factor used for the $V_{cb}$ extraction from $B\to D\ell\nu$ we used the preliminary results of Ref.~\cite{Qiu:2013ofa}. Note that now both exclusive values of $V_{cb}$ are below the inclusive determinations which disfavors right-handed currents as an explanation as we will see later in detail.

Concerning the CKM element $V_{ub}$ the situation for the different determinations is similar
\begin{align}
|V_{ub}|&=(4.41^{+0.21}_{-0.23})\times 10^{-3}\;\;({\rm inclusive})\,,\\
|V_{ub}|&=(3.40^{+0.38}_{-0.33})\times 10^{-3}\;\;(B\to \pi\ell\nu)\,,\\
|V_{ub}|&=(4.3\pm 0.7)\times 10^{-3}\;\;(B\to \tau\nu)\,,\\
|V_{ub}|&=(3.01\pm 0.57)\times 10^{-3}\;\;(B\to \rho\ell\nu)\,,
\end{align}
meaning that again the semi-leptonic exclusive determinations are below the inclusive ones. The latter agrees with the determination from $B\to\tau\nu$, which has however still quite low statistics and therefore relatively large errors \footnote{For $B\to\tau\nu$ we used the PDG average \cite{Beringer:1900zz}. While the BELLE value using leptonic tagging \cite{Hara:2010dk} agrees with the BABAR values from both the hadronic and the leptonic tag \cite{Aubert:2009wt,Lees:2012ju}, the BELLE value obtained by using hadronic tagging give a significantly lower value for $V_{ub}$. This slight tension leads to a scaling by factor of $1.3$ of the combined error \cite{Beringer:1900zz}}. The inclusive result was calculated in Ref.~\cite{Greub:2009sv} and for $B\to \rho\ell\nu$ we averaged the values of Ref.~\cite{Fu:2014pba} and multiplied the error by three in order to be conservative~\footnote{For other $B\to \rho\ell\nu$ analysis see for example~\cite{Flynn:2008zr,delAmoSanchez:2010af,Albertus:2014xwa}}. The value of $V_{ub}$ from $B\to \pi\ell\nu$ is taken from the fit of HFAG \cite{Amhis:2012bh}.

Note that the problem in the inclusive and exclusive determination of $V_{cb}$ is not directly related to the $B\to D^{(*)}\tau\nu$ problem where also a deviation from the SM of more than $3\,\sigma$ is observed by BABAR \cite{Lees:2012xj}. There, one considers the ratios ${\rm Br}[B\to D^{(*)}\tau\nu]/{\rm Br}[B\to D^{(*)}\ell\nu]$ in which the dependence on the CKM elements drops out. However, due to the heavy tau lepton involved these observables \cite{Korner:1989ve,Korner:1989qb} are sensitive to NP contributions \cite{Kamenik:2008tj,Fajfer:2012vx},  especially to charged Higgs contributions \cite{Tanaka:1994ay,Miki:2002nz,Nierste:2008qe,Kamenik:2008tj,Crivellin:2012ye} which is also true for $B\to\tau\nu$ \cite{Hou:1992sy,Akeroyd:2003zr} and the determination of $V_{ub}$ from this decay.

The question we want to address in this article is if these deviations among the different determination in the values of $V_{cb}$ and $V_{ub}$ can be explained by physics beyond the SM. For this purpose we will first study the general effect of additional effective operators (determined at the $B$ meson scale) in the next section and consider the phenomenological implication and connection to dimension-6 gauge-invariant operators in section~\ref{sec:pheno}. Finally we conclude. 

\vspace{-1mm}
\section{New physics effects in (semi-) leptonic $B$ decays}
\vspace{-1mm}

In an effective field theory approach we can parametrize the effect of NP in a model independent way. Here we consider the most general operator basis up to dimension~6 given at the $B$ meson scale. There are two different ways how NP contributions can affect the determination of the CKM elements from tree-level $B$ decays: Through additional four-fermion operators (which can be generated at tree-level) and operators which modify the $W$ coupling to quarks (via loop-effects) and therefore the charged current after integrating out the $W$ boson. 

\vspace{-1mm}
\subsection{4-fermion operators}
\vspace{-1mm}

Let us first consider the four-fermion operators which can already be generated by integrating out heavy degrees of freedom at the tree-level. Here we consider the effective Hamiltonian $H_{\rm eff}=\sum\limits_I  {C_I O_I}$ with the additional operators
\begin{equation}
\renewcommand{\arraystretch}{2.0}
	\begin{array}{l}
O^S_R=\bar \ell {P_L}\nu \bar q{P_R}b,\;\qquad O^S_L=\bar \ell {P_L}\nu \bar q{P_L}b,\\
O^T_L=\bar \ell {\sigma _{\mu \nu }}{P_L}\nu \bar q{\sigma ^{\mu \nu }}{P_L}b,
\end{array}
\end{equation}
with $q=u,c$ and assuming the absence (i.e. heaviness) of right-handed neutrinos. We postpone the discussion of a possible vector operator since these effects can also be induced by a modified $W$-$qb$ coupling. At zero recoil (i.e. maximal momentum transfer) where the CKM elements are extracted (and neglecting small lepton masses) there is no interference of scalar and tensor operators with the SM contribution both in exclusive and inclusive semi-leptonic modes and the relative importance of the operators is the same in all decay modes: the contribution to all decays from the tensor operator is simply proportional to $|C^T_L|^2$ while for $B\to D(\pi)\ell\nu$ the scalar contribution is proportional to $|C^S_R+C^S_L|^2$ and for $B\to D^*(\rho)\ell\nu$ the additional contribution scales like $|C^S_R-C^S_L|^2$ while in the inclusive decay (in the limit of vanishing lepton and charm (up) masses) we have $|C^S_R|^2+|C^S_L|^2$. Therefore, these operator cannot explain why both exclusive determinations of $V_{ub}$ and of $V_{cb}$ are below the inclusive ones. The only exception is $B\to\tau\nu$. Here it is well know that the scalar operator generated by a charged Higgs affects the branching ratio \cite{Hou:1992sy}. In fact, a charged Higgs exchange can explain the deviation from the SM in tauonic $B$ decays \cite{Crivellin:2012ye} but also the tensor operator can achieve this \cite{Sakaki:2013bfa}. We turn now to the effect of a modified $W$ couplings.

\vspace{-2mm}
\subsection{Effects of modified $W$ couplings}
\vspace{-1mm}

The impact of NP via higher dimensional operators modifying the $W$-$qb$ quark coupling has been calculated for the inclusive decay in Ref.~\cite{Dassinger:2008as} and for the exclusive modes in Ref.~\cite{Faller:2011nj}. Assuming again the absence of (light) right-handed neutrinos we can parametrize the NP contributions via the effective Hamiltonian
\begin{widetext}
\begin{equation}
{H_{eff}} = \frac{{4{G_F}{V_{qb}}}}{{\sqrt 2 }}\bar \ell {\gamma ^\mu }{P_L}\nu \left( {(1+c_L^{qb})\bar q{\gamma _\mu }{P_L}b + g_L^{qb}\bar qi{{\mathord{\buildrel{\lower3pt\hbox{$\scriptscriptstyle\leftrightarrow$}} 
\over D} }_\mu }{P_L}b + d_L^{qb}i{\partial ^\nu }\left( {\bar q{i\sigma _{\mu \nu }}{P_L}b} \right) + L \to R} \right)\,,
\end{equation}
\end{widetext}
where $q=u,c$ and the Wilson coefficients include only the effect of NP and $D_{\mu}$ is the QCD covariant derivative. Using the results of \cite{Faller:2011nj} we find the following (approximate) NP contribution to the determination of $V_{cb}$ and $V_{ub}$ from exclusive semi-leptonic modes:

\begin{widetext}
\begin{align}
V_{cb}&=\frac{V_{cb}^{\rm SM}}{1+c_L^{cb}+c_R^{cb}-1.6{\rm GeV} (d_R^{cb}+d_L^{cb})+5.5{\rm GeV} (g_R^{cb}+g_L^{cb})} \;\;(B\to D\ell\nu)\,,\\
V_{cb}&=\frac{V_{cb}^{\rm SM}}{1+c_L^{cb}-c_R^{cb}+3.3{\rm GeV} (d_R^{cb}-d_L^{cb})} \;\;(B\to D^*\ell\nu)\,,\\
V_{ub}&=\frac{V_{ub}^{\rm SM}}{1+c_L^{ub}+c_R^{ub}-4.9{\rm GeV} (d_R^{ub}+d_L^{ub})+5.5{\rm GeV} (g_R^{ub}+g_L^{ub})} \;\;(B\to \pi\ell\nu)\,,
\end{align}

\begin{figure*}
\includegraphics[width=0.48\textwidth]{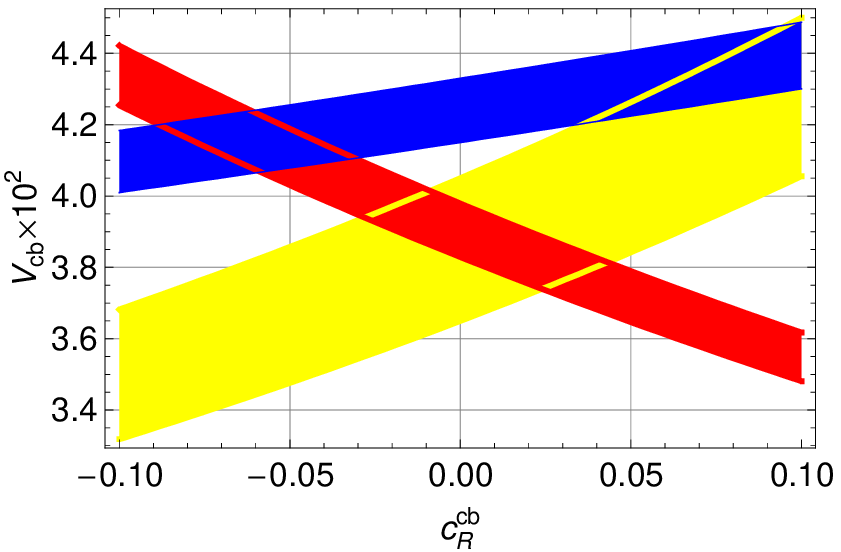}
~~~
\includegraphics[width=0.46\textwidth]{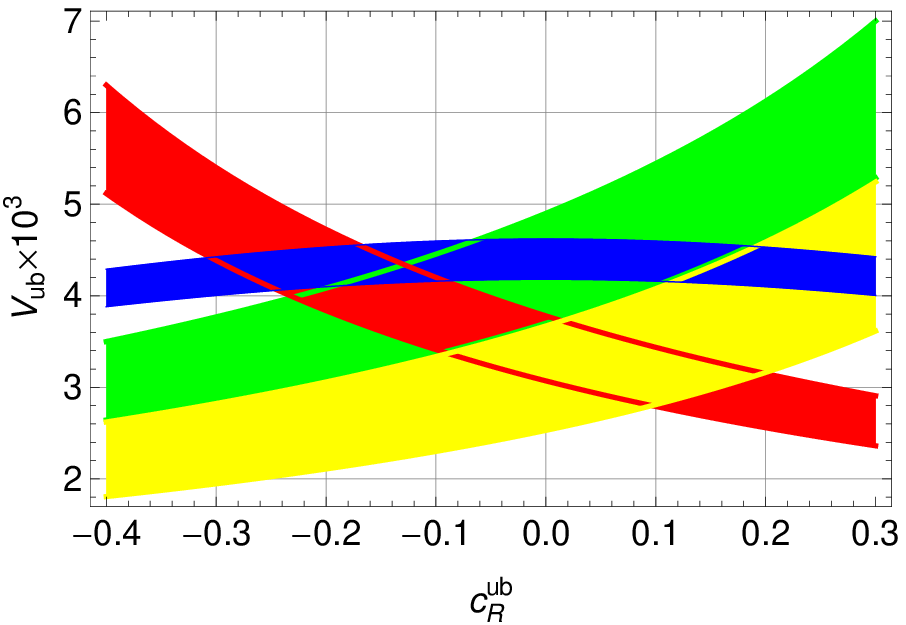}
\caption{\label{VR} Left: $\left|V_{cb}\right|$ as a function of
  $c_R^{cb}$ extracted from different
  processes. Blue(darkest): inclusive decays, red: $B\to D^*\ell\nu$, yellow: $B\to D \ell\nu$. We assumed that $c_R^{cb}$ is real. Note that with the current data, there is no point in parameter space bringing all different determination into agreement.
	Right: $\left|V_{ub}\right|$ as a function of
  $c_R^{ub}$ extracted from different
  processes. Blue: inclusive decays, red(gray): $B\to \pi\ell\nu$,
  yellow: $B\to \rho \ell\nu$, green: $B\to \tau\nu$. $c_R^{ub}$ is assumed to be real.}
\end{figure*}

\begin{align}
V_{ub}&=\frac{V_{ub}^{\rm SM}}{1+c_L^{ub}-c_R^{ub}+4.5{\rm GeV} (d_R^{ub}-d_L^{ub})} \;\;(B\to \rho\ell\nu)\,,
\label{VubVcb}
\end{align}
\end{widetext}

For the inclusive determination of $V_{cb}$ the NP effects were calculated in Ref.~\cite{Dassinger:2008as}:
\begin{equation}
V_{cb}=\frac{V_{cb}^{\rm SM}}{1+c_L^{cb}-0.34 c_R^{cb}-0.03{\rm GeV} d_R^{cb}+0.015{\rm GeV} d_L^{cb}}\,,\nonumber
\label{Vcbinclusive}
\end{equation}
For the contribution of $c_R^{cb}$ in \eq{Vcbinclusive} we used the result of Ref.~\cite{Feger:2010qc} where a global fit to all hadronic and leptonic moments was performed. We also used the result of Ref.~\cite{Feger:2010qc} to estimate the total impact of $d_{L,R}^{ub}$ and $g_{L,R}^{ub}$ on $V_{cb}$ given the various hadronic and leptonic moments in Ref.~\cite{Dassinger:2008as}. 
This is possible since the relative effect on the moment of $d_{L,R}^{cb}$ and $g_{L,R}^{cb}$ is very similar to $c^{cb}_R$. Since the inclusive $V_{cb}$ mode is not very sensitive to $d_{L,R}^{cb}$ and $g_{L,R}^{cb}$ this approximation suffices for our purpose. Concerning the inclusive determination of $V_{ub}$ the impact of NP is expected to be even smaller because of the much smaller up-quark mass and we neglect this effect.
Concerning $B\to\tau\nu$ the quantities $d_R^{ub}$ and $d_L^{ub}$ have an important effect:
\begin{equation}
V_{ub}=\frac{{V_{ub}^{\rm SM}}}{{1 + \left( {\frac{{m_B^2 - m_b^2}}{{{m_b}}}} \right)\left( {d_R^{ub} - d_L^{ub}} \right)}}\;\;(B\to \tau\nu)\,.
\end{equation}

\section{Phenomenological analysis and results}
\label{sec:pheno}

We are now in a position to examine whether the difference between the different determinations on the CKM elements can be due to NP effects. As noted in the last section, four fermion operators cannot bring the inclusive and exclusive determinations into agreement so we only consider the effect of a modified $W$ coupling here. First note that any NP contained in $c_L$ only amounts to an overall scaling of all CKM elements. The simplest possibility to explain differences between the inclusive and exclusive determinations would be  a right-handed charged currents generating $c_R$ (first studied in the context of left-right symmetric models \cite{Senjanovic:1975rk}) both for $V_{cb}$~\cite{Voloshin:1997zi,Dassinger:2007pj,He:2009hz,Crivellin:2009sd,Buras:2010pz,Faller:2011nj} and $V_{ub}$ \cite{Chen:2008se,Crivellin:2009sd,Buras:2010pz}\footnote{For a similar analysis for $V_{us}$ see Ref.~\cite{Bernard:2007cf}}. It has been shown however that in LR-symmetric models the FCNC constraints on the $W'$ mass and couplings prevent a solution of the $V_{ub}$
problem \cite{Crivellin:2011ba,Blanke:2011ry}. A sizable right-handed $W$ coupling can also be generated in the MSSM \cite{Crivellin:2009sd}. However, this is not favored anymore by the current data since all exclusive determinations are below the inclusive one. We show the effect of $c_R$ on the different determination of $V_{ub}$ and $V_{cb}$ in Fig.~\ref{VR}.

\begin{figure*}[t]
\includegraphics[width=0.47\textwidth]{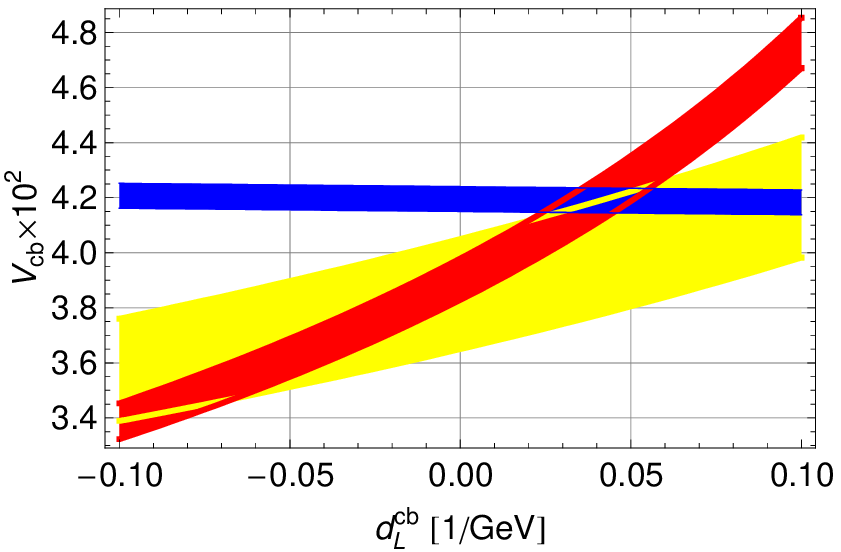}
~~~
\includegraphics[width=0.45\textwidth]{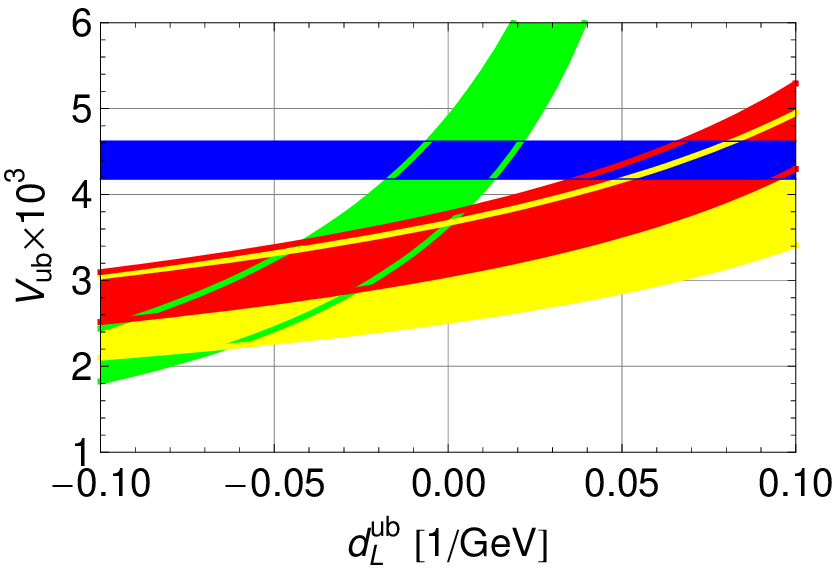}
\caption{\label{dL} Left: $\left|V_{cb}\right|$ as a function of
  $d_L^{cb}$ extracted from different
  processes. Blue: inclusive decays. Red: $B\to D^*l\nu$.
  Yellow: $B\to D \ell\nu$. Right: $|V_{ub}|$ extracted from different
  processes. Blue: inclusive decays, red(gray): $B\to \pi \ell\nu$,
  yellow: $B\to \rho \ell\nu$, green: $B\to \tau\nu$. $d_L^{qb}$ is assumed to be real.}
\end{figure*}

From \eq{VubVcb} we can see immediately, that $d_R^{qb}$ cannot bring all determinations into agreement with the current data since the effect in $B\to D\ell\nu$ and $B\to D^*\ell\nu$ ($B\to \pi\ell\nu$ and $B\to \rho\ell\nu$) is opposite. Also $g_{L,R}^{cb}$ ($g_{L,R}^{ub}$) alone is not sufficient since it affects only $B\to D^*\ell\nu$ ($B\to \rho\ell\nu$). This means that we are left with $d_L^{qb}$. The effect of $d_L^{cb}$ and $d_L^{ub}$ on the determination of $V_{cb}$ and $V_{ub}$ is shown in Fig.~\ref{dL}. We can see that for $V_{cb}$ all different determination can be brought into agreement. For $V_{ub}$ also the  exclusive semi-leptonic results can be brought into agreement with the inclusive one but a tension with $B\to\tau\nu$ is generated. 
Since $B\to\tau\nu$ is the only process under consideration involving a heavy tau lepton,  one could in principle  argue that  additional operators (most likely scalar ones for example induced by a charged Higgs) affect this decay and bring all determinations into agreement. However, as we discuss it below, it is not realistic to expect a NP contribution to be responsible for the required value of the $d_L^{qb}$.

The operators considered so far were only invariant under the gauge group $U(1)$ of the electromagnetic interactions but not under the complete SM gauge group $\rm{SU(3)_C \times SU(2)_L\times U(1)_Y}$ which any model of NP with particles above the EW scale should respect. The complete set of gauge-independent operators up to dimension-6 was derived in Ref.~\cite{Buchmuller:1985jz} and reduced to a minimal set in Ref.~\cite{GIMR}. 
Following the notation of Ref.~\cite{GIMR} the operators corresponding to $d_{L,R}$ are 
\begin{equation}
\begin{array}{l}
Q_{uW}^{ij} = 1/\Lambda^2\left( {\bar q_i{\sigma ^{\mu \nu }}u_j} \right){\tau ^I}\tilde \varphi W_{\mu \nu }^I\,,\\
Q_{dW}^{ij} = 1/\Lambda^2\left( {\bar q_i{\sigma ^{\mu \nu }}d_j} \right){\tau ^I}\varphi W_{\mu \nu }^I\,.
\end{array}
\end{equation}
First we can estimate the necessary size of the corresponding dimensionless Wilson coefficient $C_{dW}^{ij}$
\begin{equation}
	C_{dW}^{ij}\approx  g_2 V_{ij} d^{ij}_L\frac{\Lambda^2}{v}\,,
\end{equation}
which is at least of order one. Since this operator can only be induced at the loop-level, one would need non-perturbative NP interactions. Furthermore, from these expression we can see that any modification of the $W$ coupling, incorporated in $d_{L,R}$, would also lead to a modification of $Z$-quark couplings which only differs by a CKM rotation. Even for flavour-diagonal $Z$-quark couplings (in order not to violate bounds from FCNC processes) one gets a very large correction to $Z- b \bar b$ which is a factor $\cos(\theta_W)/{V_{qb}}$ larger than the contribution to the $W$ coupling. Applying the result of Ref.~\cite{Crivellin:2013hpa} to the case of bottom quarks we find the following correction to the decay width:
\begin{equation}
\Delta {\Gamma}\left[ {{Z^0} \to \bar b b } \right] \approx \frac{{{m_Z g_2^2}}}{{48\pi}}{{{\left|m_W {d_{L}^{23}} \right|}^2}}\,.
\end{equation}
For $d_{L}^{23}\approx 0.03/{\rm GeV}$, as required to explain $V_{cb}$, this is of the same order as the measured total width of the $W$ boson of approximately 2.49 GeV. Therefore, the current discrepancies between the inclusive and exclusive determinations of $V_{ub}$ and $V_{cb}$ cannot be explained by a model of NP respecting the SM gauge symmetries.

This means that the differences among the different determinations of the CKM elements must be due to experimental problems (i.e. statistical fluctuations and/or underestimated systematic errors) or due to uncertainties in the theoretical determinations of $V_{qb}$ within the SM. While the situation for $V_{cb}$ is rather clear, the conclusion for $V_{ub}$ depends crucially on $B\to \rho \ell\nu$. Indeed, if the $V_{ub}$ value extracted from this decay would be higher, a right-handed $W$ coupling could still bring the different determinations into agreement (as it is clear from fig.1). Hence, an improved determination of $V_{ub}$ from $B\to \rho \ell \nu$ as well as an analysis of right-handed currents over the full $q^2$ range would be desirable \footnote{After submission of this article, such an analysis was performed in Ref.~\cite{Bernlochner:2014ova}.}.

\section{Conclusions and Outlook}

In this article we examined if NP can explain the differences between the inclusive and exclusive determinations of $V_{ub}$ and $V_{cb}$. Using an EFT approach we found that there is only one operator capable of doing this, which corresponds to a modified momentum dependent $W$-$qb$ coupling. However, in an $SU(2)$ invariant theory of physics beyond the SM the corresponding Wilson coefficient would need to be unacceptably large, violating electroweak precision constraints on the $Z$-$bb$ coupling, ruling out a NP explanation. Therefore, the differences between the inclusive and exclusive determinations must be due to underestimated uncertainties in the theoretical and/or the experimental analysis.

Clearly, the current situation requires close reexamination of the theory predictions for all inclusive and exclusive determination of the CKM elements $V_{ub}$ and $V_{cb}$. In particular, an improved analysis of $b\rightarrow \rho \ell \nu$ would be very desirable, since $V_{ub}$ from this decay mode is crucial for (dis) favoring a right-handed $W$ coupling explanation.

Precise predictions for $V_{ub}$ are essential to judge if there is NP in $B\to\tau\nu$ and reducing the error in $V_{cb}$ is indispensable for precision predictions for $B_s\to\mu^+\mu^-$ as well as for $\epsilon_k$ and $K\to\pi\nu\nu$. In our analysis we assumed the absence of light right-handed neutrinos. Relaxing this assumption will enlarge the operator basis and would require a separate analysis.
\smallskip

{\bf Acknowledgments} {\small We thank Christoph Greub, Martin L\"uscher and Sascha Turczyk for useful discussion and Stefan Meinel for bringing the new lattice determination of the form factors from lattice QCD to our attention. We also thank Ulrich Nierste and Thomas Becher for valuable comments on the manuscript and Florian Bernlochner for bringing inconsistencies in the analysis of Ref.~\cite{Albertus:2014xwa} to our attention. AC is supported by a Marie Curie Intra-European Fellowship of the European Community's 7th Framework Programme under contract number (PIEF-GA-2012-326948). SP has been supported by the National Science Center under the research grants DEC-2011/01/M/ST2/02466, DEC-2012/04/A/ST2/00099, DEC-2012/05/B/ST2/02597. The authors are grateful to the Mainz Institute for Theoretical Physics (MITP) for its hospitality and partial support during the completion of this work.}

\bibliography{VubVcb}

\end{document}